\documentclass[10pt]{article}
\usepackage{graphicx}
\usepackage{amsmath}

\def\Title#1{\begin{center} {\Large #1 } \end{center}}
\def\Author#1{\begin{center}{ \sc #1} \end{center}}
\def\Address#1{\begin{center}{ \it #1} \end{center}}

\newcommand\pubblock{\rightline{\begin{tabular}{l} Proceedings of the Second Annual LHCP\\ \pubnumber\\
         \pubdate  \end{tabular}}}

\newenvironment{Abstract}{\begin{quotation} \begin{center}
             \large ABSTRACT \end{center}\bigskip 
      \begin{large}}{\end{large} \end{quotation}}

\newenvironment{Presented}{\begin{quotation} \begin{center} 
             PRESENTED AT\end{center}\bigskip 
      \begin{center}\begin{large}}{\end{large}\end{center} \end{quotation}}

\def\beq{\begin{equation}}
\def\eeq#1{\label{#1}\end{equation}}
\def\eeqn{\end{equation}}

\def\beqa{\begin{eqnarray}}
\def\eeqa#1{\label{#1}\end{eqnarray}}
\def\eeqan{\end{eqnarray}}

\let\bar=\overbar

\def\Dslash{\not{\hbox{\kern-4pt $D$}}}
\def\dslash{\not{\hbox{\kern-2pt $\del$}}}

\def\msb{{\bar{\ssstyle M \kern -1pt S}}}

\newcommand{\invfb}{\mathrm{\,fb}^{-1}}
\newcommand{\tev}{\mathrm{\,TeV}}
\newcommand{\gev}{\mathrm{\,GeV}}

\newcommand{\picosec}{\mathrm{\,ps}}
\newcommand{\femtosec}{\mathrm{\,fs}}
\newcommand{\pt}{p_{\mathrm{T}}}
\newcommand{\BF}{\mathcal{B}}

\usepackage{color}
\usepackage{lineno}

\newcommand\pubnumber{ }

\newcommand\pubdate{\today}

\def\affiliation{
On behalf of the LHCb collaboration, \\
Key Laboratory of Particle \& Radiation Imaging (Tsinghua University), \\
Ministry of Education; \\
Center for High Energy Physics, Department of Engineering Physics, \\
Tsinghua University, Beijing 100084, China }

\begin{document}

\large
\begin{titlepage}
\pubblock

\vfill
\Title{  Heavy flavour spectroscopy at LHC  }
\vfill

\Author{ Yiming Li  }
\Address{\affiliation}
\vfill
\begin{Abstract}

The $pp$ collision data collected in the LHC Run I provides a great opportunity for heavy flavour studies. The latest results on exotic states, heavy baryon and $B_c^+$ mesons are reviewed.

\end{Abstract}
\vfill

\begin{Presented}
The Second Annual Conference\\
 on Large Hadron Collider Physics \\
Columbia University, New York, U.S.A \\ 
June 2-7, 2014
\end{Presented}
\vfill
\end{titlepage}
\def\thefootnote{\fnsymbol{footnote}}
\setcounter{footnote}{0}
%

\normalsize 



\section{Introduction}
Thanks to the large center-of-mass energy available at the LHC, $b\bar{b}$ and
$c\bar{c}$ pairs are produced prolifically, which provides great opportunities
for studying the production and properties of heavy hadrons. 
This is not only important
itself as tests and inputs to QCD models, but also because they have to be 
well understood as the Standard Model background in the search for new physics.
The major LHC detectors are complementary to each other in the study of heavy
flavour spectroscopy by covering different acceptance and kinematic ranges:
general purpose detectors like ATLAS and CMS cover high $\pt$ and low rapidity
range, while the forward spectrometer LHCb has access to lower $\pt$ and higher 
rapidity region.
This proceeding reports recent results from these experiments, on exotic states,
heavy baryons and $B_c^+$ meson.

\section{Exotic states}
\begin{figure}[htb]
\centering
\includegraphics[height=0.2\textheight]{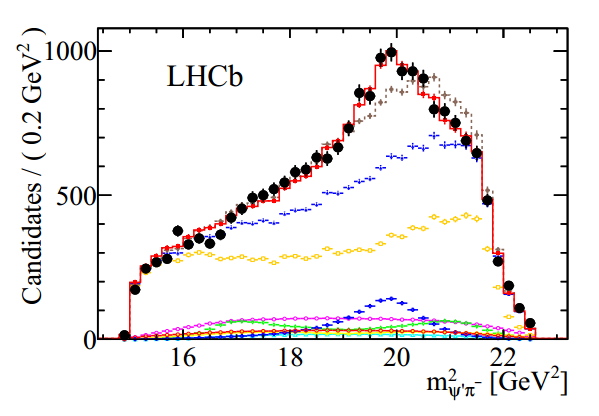}
\includegraphics[height=0.2\textheight]{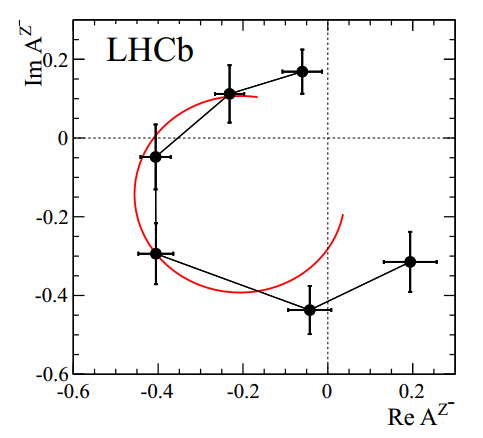}
\caption{(Left) Distribution of $m^2_{\psi(2S)\pi^-}$. Black dots are data, the red solid and brown dashed lines represent the total fit with and without the $Z(4430)^-$ component. (Right) Fitted values of $Z(4430)^-$ amplitude in six $m^2_{\psi(2S)\pi^-}$ bins shown in an Argand diagram.}
\label{fig:Exotics-Z4430}
\end{figure}

The exotic state that attracts a lot of attention recently
is the charged charmonium-like $Z(4430)^-$.
This state was first reported as a $\psi(2S)\pi^{-}$ bump in 
$B^0 \to \psi(2S) \pi^- K^+$ by Belle collaboration in 2008; BaBar collaboration
could explain the enhancement as a reflection of the known $K^*$ states, 
but did not rule out the existence of $Z(4430)^-$ either. 
With a $B^0$ signal yield an order of magnitude larger than BaBar or Belle
detectors have, LHCb collaboration performs a full amplitude analysis
considering known $K^*$ states, and observe the $Z(4430)^{-}$ with a
significance larger than $13.9 \sigma$, as shown in Figure \ref{fig:Exotics-Z4430}~\cite{Aaij:2014jqa}. The Argand diagram of the $Z(4430)^-$ amplitude (Figure
\ref{fig:Exotics-Z4430}) shows the resonance behaviour for the first time.
The spin-parity is measured to be $1^+$, by excluding 
$0^-, 1^-, 2^-, 2^+$ hypotheses by at least $9.7\sigma$. 
For a charged charmonium state, $Z(4430)^-$ has a minimum quark content of 
$c\bar{c}d\bar{u}$, which clearly does not fit into traditional quark model.

\begin{figure}[htb]
\centering
\includegraphics[width=0.45\textwidth]{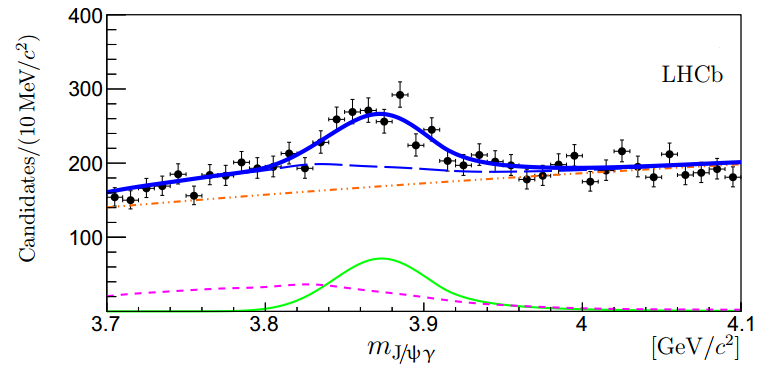}
\includegraphics[width=0.45\textwidth]{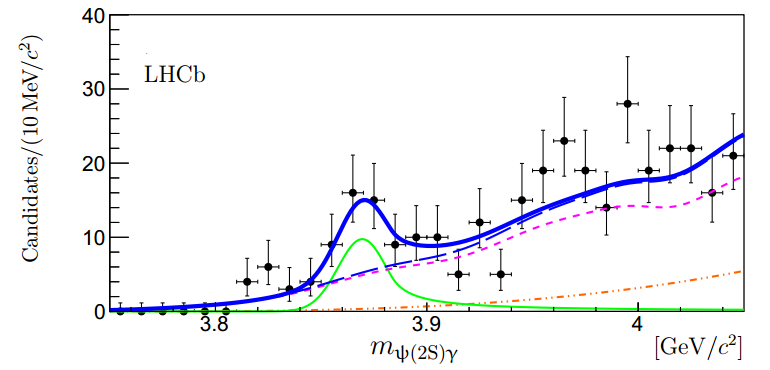}
\caption{Invariant mass distributions of (left) $J/\psi\gamma$ and (right) $\psi(2S)\gamma$ from $B$ decays.}
\label{fig:Exotics-X3872}
\end{figure}

$X(3872)$ is the first charmonium-like exotic state ever observed. Its
quantum number is finally pinned down to $1^{++}$ in 2013~\cite{Abulencia:2006ma,Aaij:2013zoa}, but the nature
is still unclear, drawing a lot of theoretical interests.
Useful information can be obtained from its radiative decay since various 
interpretations predict very different values for the ratio
$R_{\psi\gamma} \equiv \BF(X(3872) \to \psi(2S)\gamma)/\BF(X(3872) \to J/\psi\gamma)$.
LHCb lately finds a $4.4\sigma$ evidence of $X(3872) \to \psi(2S)\gamma$ decay
in $B^+ \to X(3872)K^+$~\cite{Aaij:2014ala} (as shown in 
Figure \ref{fig:Exotics-X3872}), and measured its branching fraction
relative to $X(3872) \to J/\psi\gamma$: $R_{\psi\gamma} = 2.46 \pm 0.64 \pm 0.29$, 
where the first uncertainty is statistical and the second systematic, 
as followed in the rest of the proceeding. 
This result is consistent with expectations of a charmonium $c\bar{c}(2^3P_1)$ 
or a molecule-charmonium mixture interpretation, but does not support 
a pure $D\bar{D}^*$ molecule interpretation.

\begin{figure}[htb]
\centering
\includegraphics[width=0.55\textwidth]{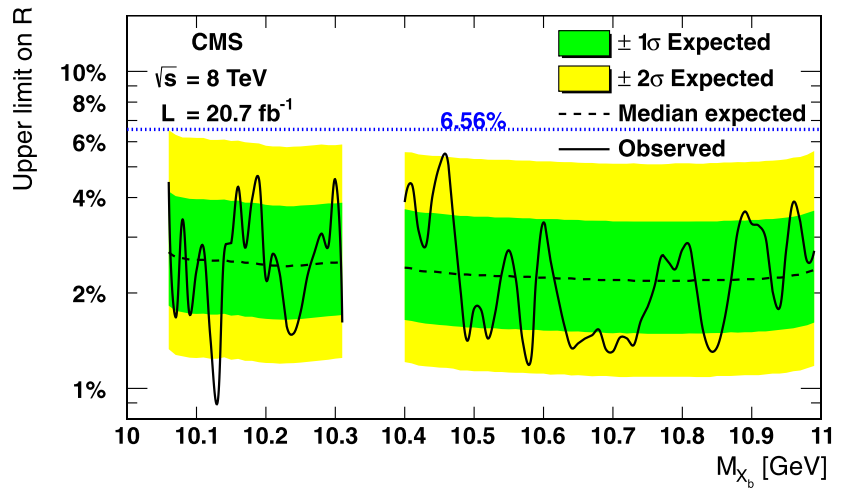}
\caption{Upper limit at the 95\% confidence level on $R_{X_b}$ (see definition in text) as a function of $X_b$ mass.}
\label{fig:Exotics-Xb}
\end{figure}

A search for the bottomonium counterpart of $X(3872)$ is recently 
performed in $X_b \to \Upsilon(1S)\pi^+\pi^-, \Upsilon(1S) \to \mu^+\mu^-$ 
decay by CMS \cite{Chatrchyan:2013mea}. 
No evidence of $X_b$ is observed, and upper limits are set
at 95\% confidence level on the ratio $R_{X_b} \equiv \sigma(pp \to X_b \to \Upsilon(1S)\pi^+\pi^-)/\sigma(pp \to \Upsilon(2S) \to \Upsilon(1S)\pi^+\pi^-)$ 
as a function of $X_b$ mass (Figure~\ref{fig:Exotics-Xb}).
The upper limits are in the range of $0.9-5.4\%$ for $X_b$ mass between
10 and 11 GeV.

\section{Heavy baryons}
\begin{figure}[htb]
\centering
\includegraphics[height=0.16\textheight]{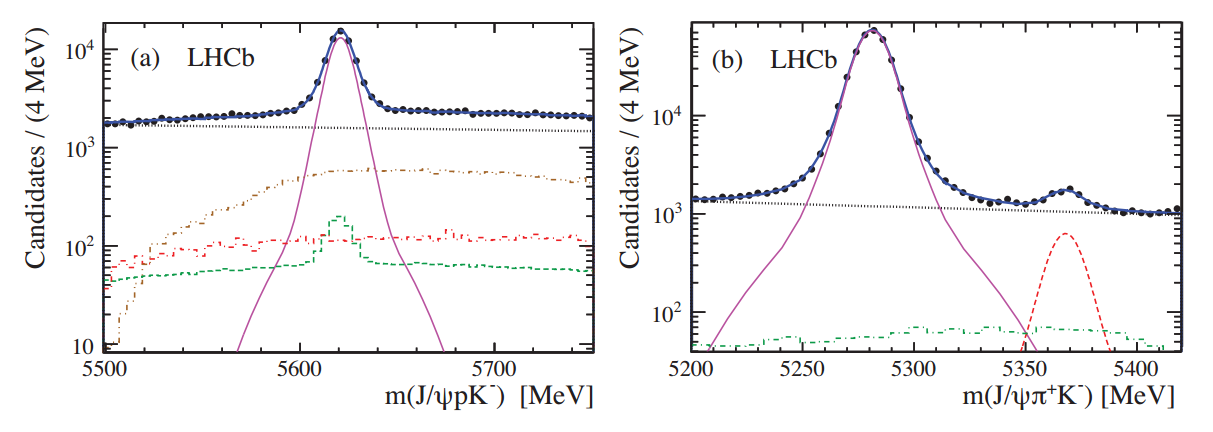}
\includegraphics[height=0.17\textheight]{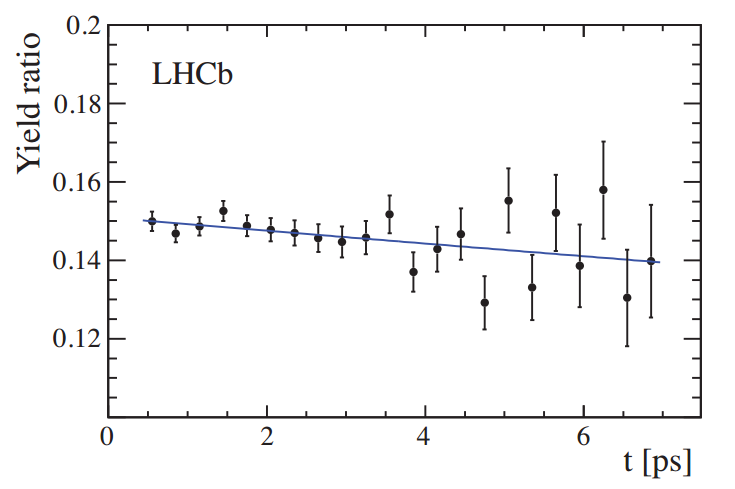}
\caption{Fits to invariant mass distribution of (left) $J/\psi p K^-$ and (middle) $J/\psi \pi^+ K^-$ combinations. The solid purple lines represents the fitted $\Lambda_b^0$ and $\bar{B}^0$ signals. (Right) The yield ratio of $N(\Lambda_b^0)/N(\bar{B}^0)$ as a function of decay time.}
\label{fig:Baryon-Lb}
\end{figure}

According to the Heavy Quark Expansion (HQE) theory, 
the $b$ baryon lifetime is expected to be close to that of the $B$ mesons, 
and $\tau(\Lambda_b^0)/\tau(\bar{B}^0)$ should
differ from unity by no more than a few percent. However LEP results indicate
this ratio is much smaller, which became a puzzle for the last decade.
With $pp$ collision data at $\sqrt{s} = 7 \tev$, ATLAS~\cite{Aad:2012bpa} 
and CMS~\cite{Chatrchyan:2013sxa} determined $\Lambda_b^0$ lifetime, 
resulting in a ratio over $B^0$ lifetime much closer to one with large 
uncertainties. LHCb precisely measures $\tau(\Lambda_b^0)/\tau(\bar{B}^0)$
with $1 \invfb$ data~\cite{Aaij:2013oha} and the result is consistent with HQE prediction.
Lately LHCb updated this measurement with $3 \invfb$ full data from Run~I~\cite{Aaij:2014zyy}. 
Figure \ref{fig:Baryon-Lb} shows the reconstructed $\Lambda_b^0$ and $\bar{B}^0$
signals using similar decay final states $\Lambda_b^0 \to J/\psi p K^-$
and $\bar{B}^0 \to J/\psi \bar{K}(892)^{*0} (\bar{K}(892)^{*0} \to \pi^+ K^-)$, 
as well as their yield ratio as function of decay time. The lifetime ratio is 
measured to be $\tau(\Lambda_b^0)/\tau(\bar{B}^0) = 0.974 \pm 0.006 \pm 0.004$.
This agrees with previous LHCb result and HQE prediction, and gives
the most precise $\Lambda_b^0$ lifetime measurement using $\tau(\bar{B}^0)$
world average.
In another measurement of $\tau(\Lambda_b^0)$, LHCb uses a different decay
channel of $\Lambda_b^0 \to J/\psi \Lambda$ with $1 \invfb$~\cite{Aaij:2014owa}. 
Combined result gives $\tau(\Lambda_b^0) = 1.468 \pm 0.009 \pm 0.008 \picosec$.
The latter analysis also gives the most precise single measurement of 
$B^+$, $B^0$ and $B_s^0$ (effective) lifetime, as listed in Table \ref{tab:Baryon-Lb}.

\begin{table}[t]
\begin{center}
\begin{tabular}{lc}  
\hline Lifetime & Value (ps) \\
\hline
$\tau_{B^+ \to J/\psi K^+}$   & 1.637 $\pm$ 0.004 $\pm$ 0.003 \\
$\tau_{B^0 \to J/\psi K^{*0}}$ & 1.524 $\pm$ 0.006 $\pm$ 0.004 \\
$\tau_{B^0 \to J/\psi K_{\mathrm{S}}^0}$ & 1.499 $\pm$ 0.013 $\pm$ 0.005 \\
$\tau_{\Lambda_b^0 \to J/\psi \Lambda}$ & 1.415 $\pm$ 0.027 $\pm$ 0.006 \\
$\tau_{B_s^0 \to J/\psi \phi}$ & 1.480 $\pm$ 0.011 $\pm$ 0.005 \\
\hline
\end{tabular}
\caption{Measured $B^+$, $B^0$, $\Lambda_b^0$ and $B_s^0$ (effective) lifetime as in Ref.~\cite{Aaij:2014owa}.}
\label{tab:Baryon-Lb}
\end{center}
\end{table}

\begin{figure}[htb]
\centering
\includegraphics[height=0.36\textheight]{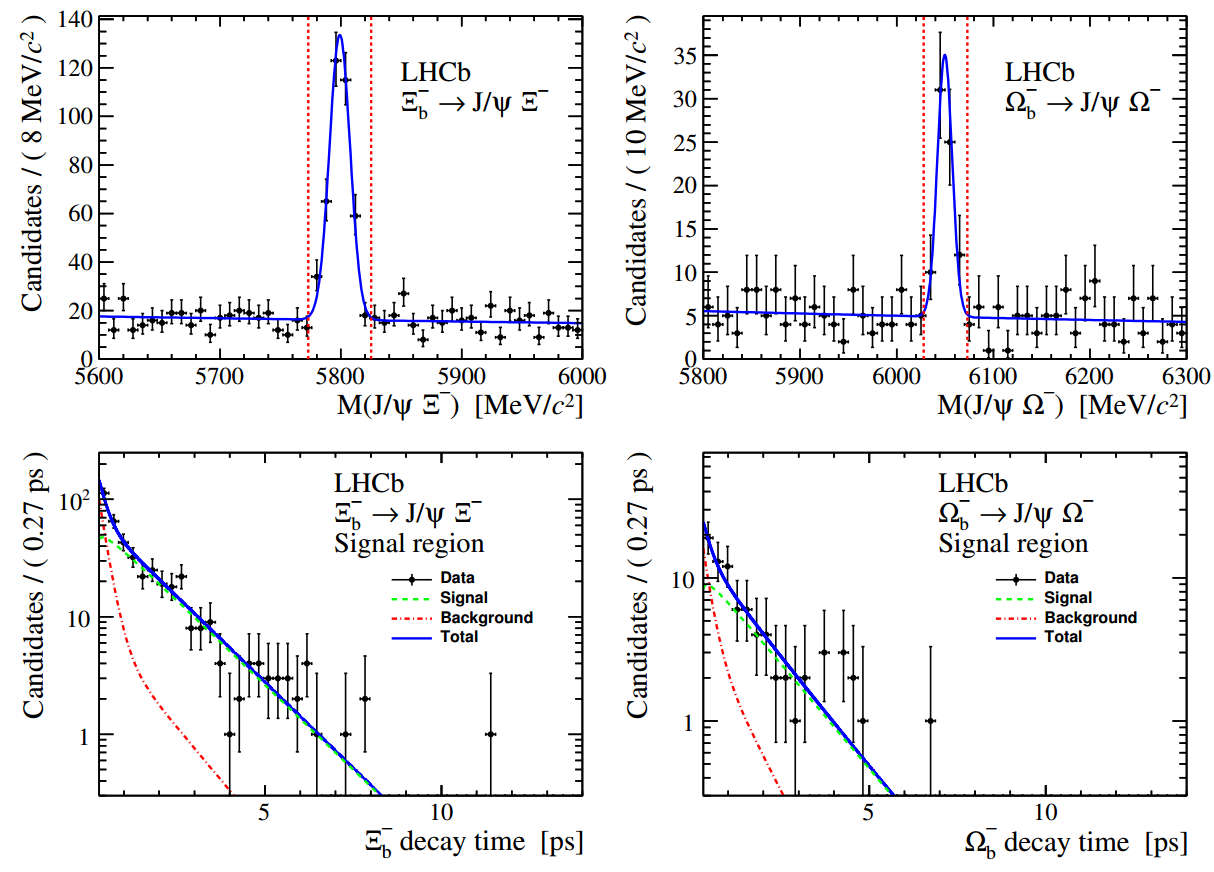}
\caption{(Top) Invariant mass of (left) $J/\psi\Xi^-$ and (right) $J/\psi\Omega^-$ combinations. (Bottom) Decay time of reconstructed (left) $\Xi_b^-$ and (right) $\Omega_b^-$ candidates.}
\label{fig:Baryon-Xib}
\end{figure}

Unlike the $\Lambda_b^0$ state, 
the bottom strange baryons such as $\Xi_b$ and $\Omega_b$ are
less abundantly produced, hence less studied. 
LHCb lately measured their lifetime, using $\Xi_b^- \to J/\psi\Xi^-$ and 
$\Omega_b^- \to J/\psi\Omega^-$ channels, with subsequent decays of
$\Xi^- \to \Lambda\pi^-$, $\Omega^- \to \Lambda K^-$, $\Lambda \to p \pi^-$
and $J/\psi \to \mu^+\mu^-$~\cite{Aaij:2014sia}. 
The reconstructed mass and decay time of 
$\Xi_b^-$ and $\Omega_b^-$ are shown in Figure \ref{fig:Baryon-Xib}, 
their lifetime are measured to be 
$\tau(\Xi_b^-) = 1.55 _{-0.09}^{+0.10} \pm 0.03 \picosec$ and
$\tau(\Omega_b^-) = 1.54 _{-0.21}^{+0.26} \pm 0.05 \picosec$. These are 
the most precise measurements to date, consistent with CDF results~\cite{Aaltonen:2009ny,Aaltonen:2014wfa} and with theoretical predictions. 

\section{$B_c^+$ physics}
\begin{figure}[htb]
\centering
\includegraphics[height=0.18\textheight]{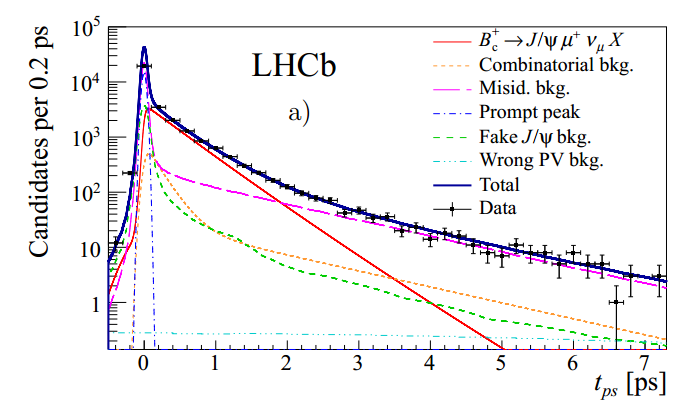}
\includegraphics[height=0.18\textheight]{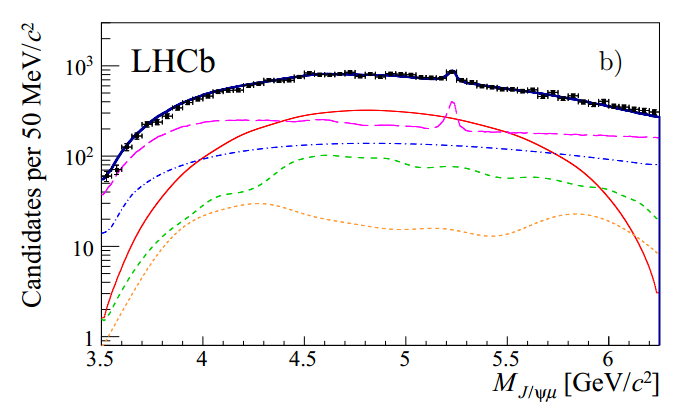}
\caption{(Left) Pseudo decaytime of reconstructed $B_c^+$ candidates; (right) invariant mass distributions of $J/\psi\mu$ combinations.}
\label{fig:Bc-lifetime}
\end{figure}

The $B_c^+$ state is the ground state of a family of unique mesons 
that consist of two different heavy flavour quarks. 
Its production cross-section at the LHC is expected to be an order of magnitude 
larger than it was at Tevatron (where it was discovered), 
and this allows more detailed study on its properties.
Using the final states of $B_c^+ \to J/\psi\pi^+$, the ratio $R_{\sigma} = \sigma(B_c^+)\BF(B_c^+ \to J/\psi\pi^+)/\sigma(B^+)\BF(B^+ \to J/\psi K^+)$ is measured by LHCb~\cite{Aaij:2012dd} and CMS~\cite{CMS:2013pua}
 in different kinematic regions as shown in Table \ref{tab:Bc-production}.
The dominating systematic uncertainties for $R_{\sigma}$ measurements come from
the uncertainty of $B_c^+$ lifetime which was measured only at Tevatron. 
However this situation has been changed lately, since LHCb measured the lifetime
with better precision: $\tau(B_c^+) = 509 \pm 8 \pm 12 \femtosec$~\cite{Aaij:2014bva} using semileptonic decay $B_c^+ \to J/\psi\mu^+\nu$.
Figure~\ref{fig:Bc-lifetime} shows the results of simultaneous fits to 
the distributions of reconstructed $B_c^+$ pseudo decaytime and invariant mass
of the $J/\psi\mu^+$ combinations. The pseudo decaytime is defined as
$\boldsymbol{\mathit{p}}\cdot(\boldsymbol{\mathit{v-x}})M_{3\mu}/|\boldsymbol{\mathit{p}}|^2$, where
$\boldsymbol{p}$ is the three-momentum of the $J/\psi\mu$ system 
in the laboratory frame, and $\boldsymbol{v}$ and $\boldsymbol{x}$ are the measured
position of the $B_c^+$ decay and production vertices.
This improvement will benefit
many other $B_c^+$ measurements such as its mass and decay branching ratios.
\begin{table}[t]
\begin{center}
\begin{tabular}{ccc}  
\hline Experiment & $R_{\sigma} = \frac{\sigma(B_c^+)\BF(B_c^+ \to J/\psi\pi^+)}{\sigma(B^+)\BF(B^+ \to J/\psi K^+)}$ & Kinematic range \\
\hline
LHCb~\cite{Aaij:2012dd} & $(0.68 \pm 0.10 \pm 0.03 \pm 0.05)\%$ & 
$\pt > 4 \gev$, $2.5 < \eta < 4.5$ \\
CMS~\cite{CMS:2013pua}  & $(0.48 \pm 0.05 \pm 0.04 \,_{-0.03}^{+0.05})\%$ &
$\pt > 15 \gev$, $|y| < 1.6$ \\
\hline
\end{tabular}
\caption{$R_{\sigma}$ measured at different kinematic ranges, where the uncertainties are statistical, systematic and that caused by uncertainty on $B_c^+$ lifetime using world average.}
\label{tab:Bc-production}
\end{center}
\end{table}

\begin{figure}[htb]
\centering
\includegraphics[height=0.2\textheight]{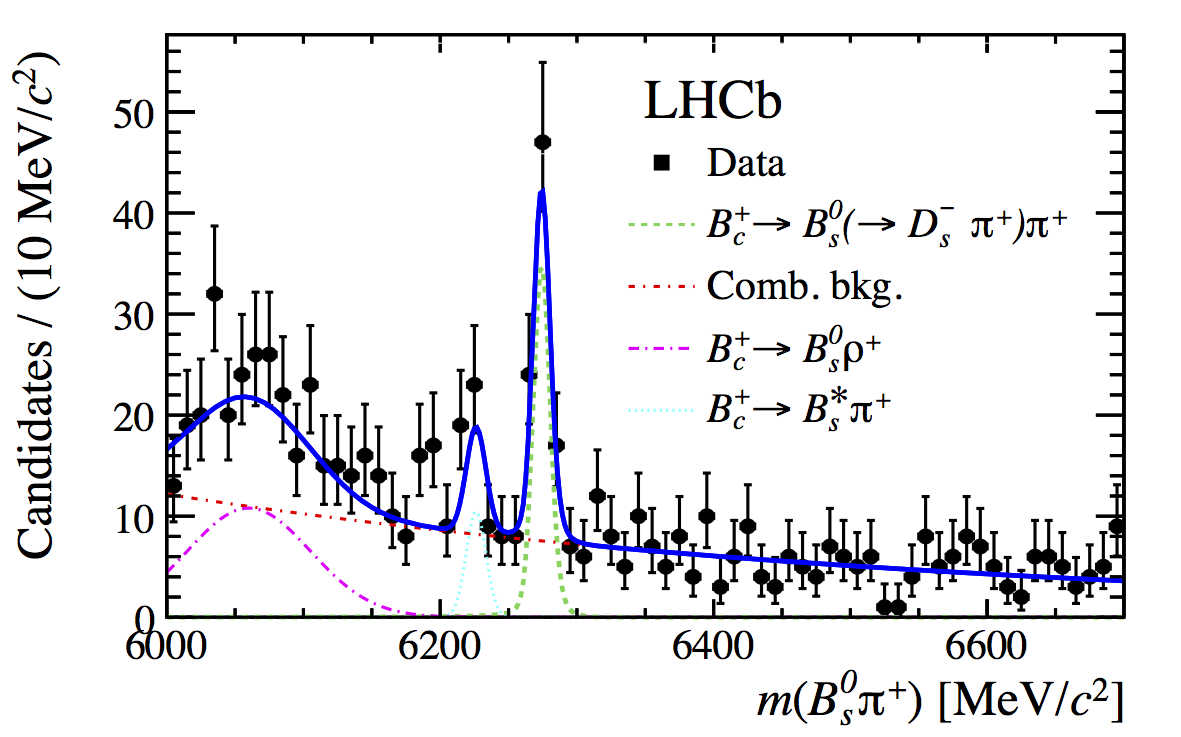}
\includegraphics[height=0.2\textheight]{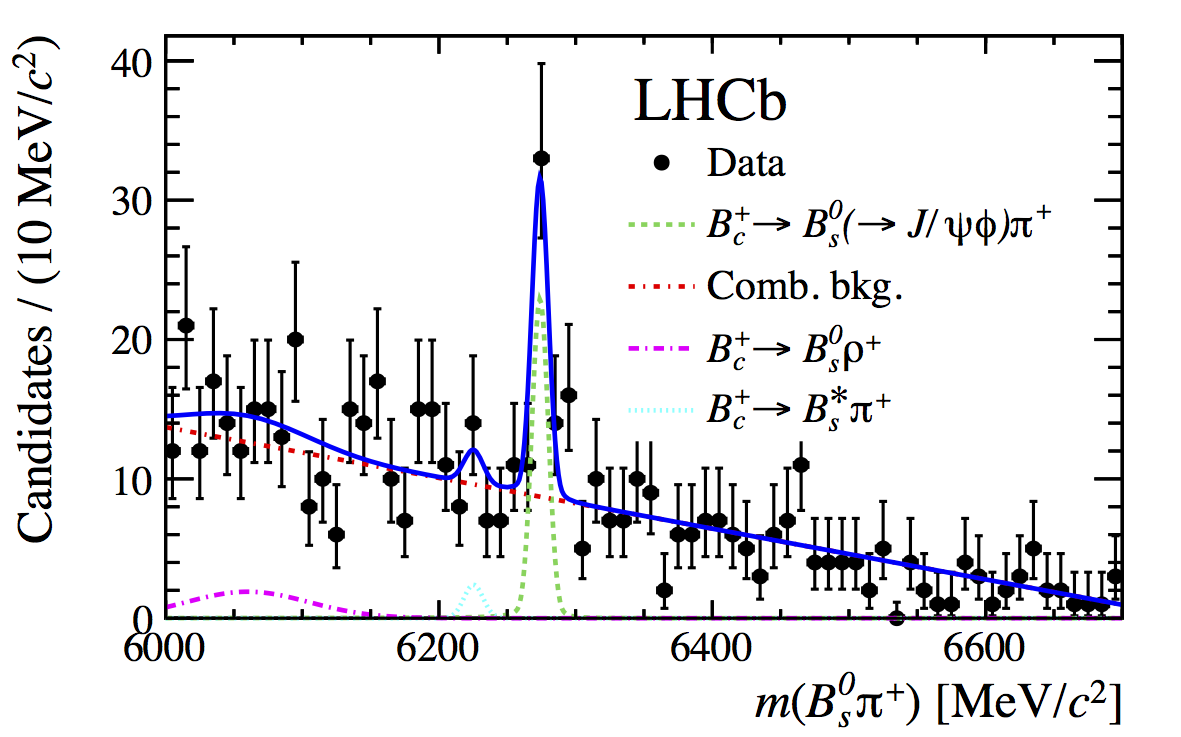}
\caption{Invariant mass of $B_s\pi^+$ combination, using (left) $B_s^0 \to D_s^-\pi^+$ and (right) $B_s^0 \to J/\psi\phi$ final states respectively.}
\label{fig:Bc-Bspi}
\end{figure}

\begin{figure}[htb]
\centering
\includegraphics[height=0.18\textheight]{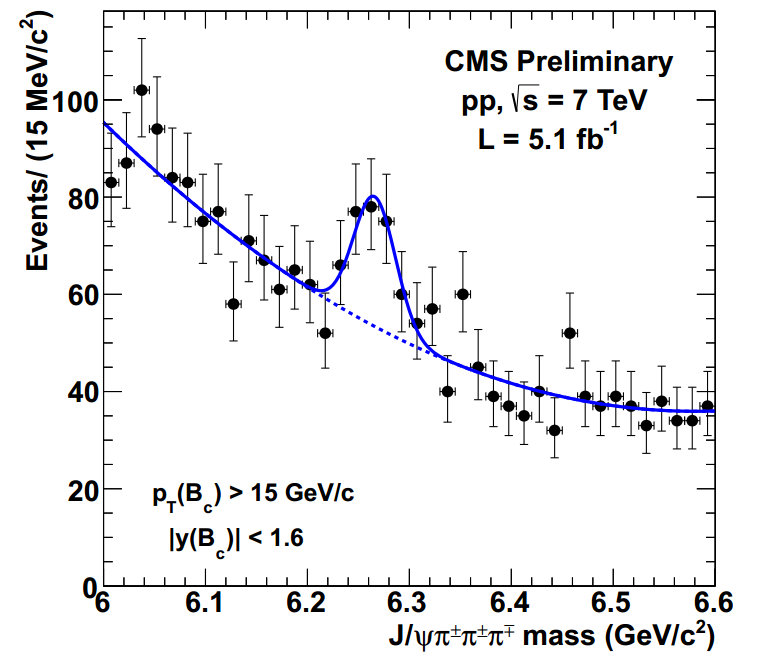}
\includegraphics[height=0.18\textheight]{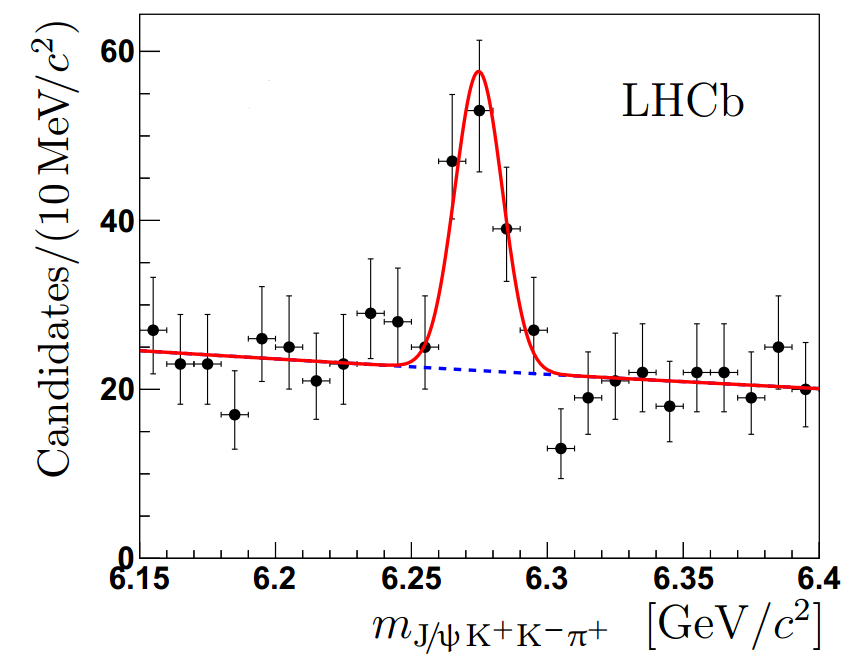}
\includegraphics[height=0.18\textheight]{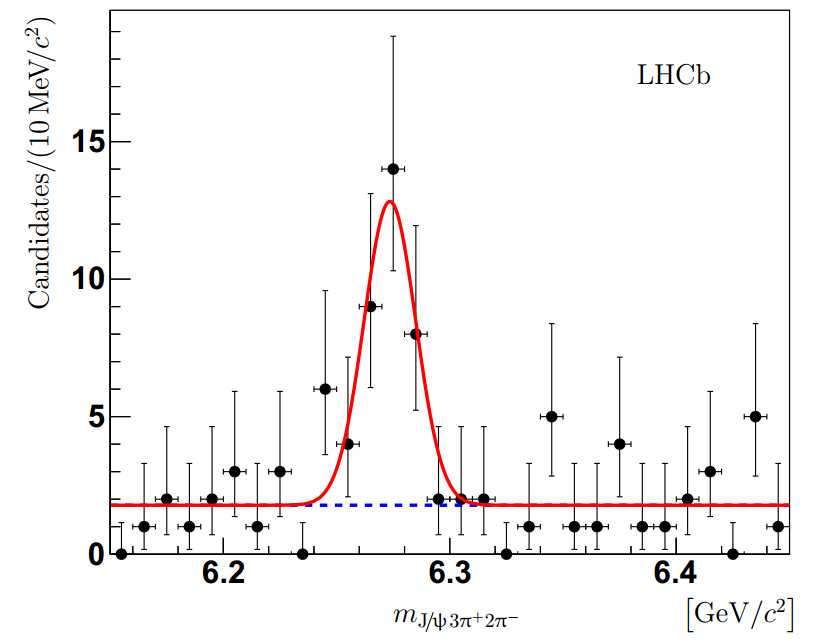}
\caption{Mass distributions of the $B_c^+$ candidates reconstructed from (left) $J/\psi\pi^+\pi^-\pi^+$, (middle) $J/\psi K^+ K^- \pi^+$ and (right) $J/\psi\pi^+\pi^+\pi^+\pi^-\pi^-$ decays.}
\label{fig:Bc-JpsiNpi}
\end{figure}

The $B_c^+$ mesons are expected to decay via many modes: either $c$ or $\bar{b}$ 
quark decays weakly with the other as spectator, or they can annihilate.
Only very few channels are experimentally observed before the operation of LHC 
due to the small production cross-section available. 
Nowadays the list of observed decays has been expanded, mostly by LHCb,
including observation of the very first $c$ quark decay $B_c^+ \to B_s^0\pi^+$~\cite{Aaij:2013cda},
as shown in Figure \ref{fig:Bc-Bspi}.
For $\bar{b}$ quark decay the list of new channels is longer, the latest being 
observation of $B_c^+ \to J/\psi K^+ K^- \pi^+$~\cite{Aaij:2013gxa} and a 4.5
$\sigma$ evidence of $B_c^+ \to J/\psi 3\pi^+ 2\pi^-$ decay~\cite{Aaij:2014bla}
(as shown in Figure \ref{fig:Bc-JpsiNpi}). 
CMS recently observed $B_c^+ \to J/\psi \pi^+\pi^-\pi^+$~\cite{CMS:2013pua}
and measured its branching fraction relative to $B_c^+ \to J/\psi\pi^+$,
the result consistent with LHCb result~\cite{LHCb:2012ag}.

\section{Conclusions}
The LHC experiments have been fruitful at heavy flavour spectroscopy studies.
This proceeding reviews some latest highlights, including observation
of a charged charmonium-like state $Z(4430)^-$, study of the $X(3872)$ radiative
decays, more precise determination of $b$-baryon lifetimes, 
and a better understanding of the $B_c^+$ properties.
As the analysis of Run I data still ongoing and Run II at higher
center-of-mass energy is starting soon, more interesting results will
keep coming out to gain us more knowledge on the heavy hadron spectroscopy.




\end{document}